\documentclass[preprint,authoryear,12pt]{elsarticle}

\usepackage{epsfig}

\usepackage{amssymb}

\usepackage[ps2pdf,%
a4paper=true,%
breaklinks=true,%
colorlinks=true,%
pdfauthor={First Author et al.},%
pdftitle={Template for manuscripts in Planetary and Space Science}%
]{hyperref}

\journal{Advances in Space Research}

\begin{document}

%%%%%%%%%%%%%%%%%%%%%%%%%%%%%%%%%%%%%%%%%%%%%%%%%%%%%%%%%%%%%%%%%%%%%%%%%%%%%
%% Frontmatter
\begin{frontmatter}


% Use the tnoteref command within \title and fnref within \author or \address for footnotes;
% use the corref command within \author for corresponding author footnotes;
% \author{Name\corref{cor1}\fnref{label2}}
% \address{Address\fnref{label3}}The Sloan Digital Sky Survey Photometric System

\title{An Unprecedented Constraint on Water Content in the Sunlit Lunar Exosphere Seen by Lunar-Based Ultraviolet Telescope of Chang'e-3 Mission}
%\tnoteref{footnote1}}
%\tnotetext[footnote1]{This template can be used for all publications in Advances in Space Research.}

% Use optional labels to link authors explicitly to addresses:

\author{J. Wang\corref{cor}}%\fnref{footnote2}}
\ead{wj@bao.ac.cn}
\author{C. Wu}
\author{Y. L. Qiu}
\author{X. M. Meng}
\author{H. B. Cai}
\author{L. Cao}
\author{J. S. Deng}
\author{X. H. Han}
\author{J. Y. Wei}
\address{National Astronomical Observatories, Chinese Academy of Sciences,
20A, Datun Road, Chaoyang District, Beijing, China, 100012}
\cortext[cor]{Corresponding author}
%\fntext[footnote2]{Additional information regarding the corresponding author}

% Url can be given like this:
% \ead[url]{http://www.elsevier.com/wps/find/authorsview.authors/latex}

%\address{National Astronomical Observatories, Chinese Academy of Sciences,
%20A, Datun Road, Chaoyang District, Beijing, China, 100012}
%\ead{jsdeng@bao.ac.cn}

%\fntext[footnote3]{Additional information about the second and third authors}
%\ead{more@email.addresses}

%\author{More Authors\fnref{footnote4}}
%\address{Address of the co-authors}
%\fntext[footnote4]{Additional information about the co-authors}
%\ead{more@email.addresses}

\begin{abstract}
%The near-ultraviolet sky background emission from the solar-forced lunar exosphere is predicted basing upon the
%current knowledge of the exosphere. The emission is calculated by a sum of the contributions from
%Rayleigh scattering, resonance fluorescence and emissive photodissociation mechanisms.
%The Moon has a tenuous exosphere, but its content is still a mystery. 
The content of $\mathrm{OH/H_2O}$ molecules in the tenuous exosphere of the Moon is still an open
issue at present.  
%However, the reported measurements and/or upper limits of the OH radicals are 2-6 orders of magnitude larger
%than all the proposed models. 
We here report an unprecedented upper limit  of the content of the OH radicals, which is obtained from the 
in-situ measurements carried out \rm by the Lunar-based Ultraviolet Telescope, a payload of Chinese Chang'e-3 mission.
 By analyzing the diffuse background in the images taken by the telescope, 
the column density 
and surface concentration of the OH radicals are inferred to be $<10^{11}\ \mathrm{cm^{-2}}$ and $<10^{4}\ \mathrm{cm^{-3}}$ 
(by assuming a hydrostatic equilibrium with a scale height of 100km), respectively, by
assuming that the recorded background is fully contributed by their resonance fluorescence emission. The resulted concentration is
lower than the previously reported value by about 
two orders of magnitude, and  is close to \rm 
the prediction of the sputtering model. In addition, the same measurements and method allow us to derive a 
surface concentration of $<10^{2}\ \mathrm{cm^{-3}}$ for the neutral magnesium, which is  
lower than the previously reported upper limit by about two orders of 
magnitude. \rm These results are the best
known of the OH (MgI) content in the lunar exosphere to date.
\end{abstract}


\end{frontmatter}

\parindent=0.5 cm

%%%%%%%%%%%%%%%%%%%%%%%%%%%%%%%%%%%%%%%%%%%%%%%%%%%%%%%%%%%%%%%%%%%%%%%%%%%%%
%% Main text
\section{Introduction}

Water is important not only for planetary evolution, but also for the existence of life. The
Moon has been believed to be anhydrous for a long time based on the returned lunar samples 
acquired by the Apollo and Luna missions. This point of view should be, however, significantly
modified at present due to the explorations carried out by the Chandrayaan-1 and Cassini missions.
Both Moon Mineralogy Mapper ($\mathrm{M^3}$), a NASA instrument onboard the Chandrayaan-1, and Visual and Infrared Mapping
Spectrometer (VIMS) on the Cassini recently reported  a wide distribution
of water on the sunlit side of the Moon
through the broad spectral absorption features at 2.8 and 3$\mu$m (Sunshine et al., 2009; Clark et al., 2009; Pieters et al., 2009). 
These features are attributed to the OH and $\mathrm{H_2O}$ molecules chemically adsorbed in silicate rocks. \rm
Although these adsorbed $\mathrm{H_2O}$/OH molecules are stable below a temperature of $\sim500$K (e.g., Hibbitts et al., 2010; Dyar et al., 2010), 
they could be released to the lunar exosphere by micro-metor impact, photon stimulated desorption
and sputtering by the high energy protons in the solar wind (e.g., Stern, 1999; Morgan et al., 1997; Wurz et al., 2007; 
Killen \& Ip, 1999; Hunten et al., 1997; Johnson et al., 1991; Killen et al., 1999; Mendillo et al., 1999).

It was known for a long time that the Moon has a tenuous exosphere (Stern, 1999). The content and distribution
of the exosphere is rather poorly understood at present because of its extremely
low density, 
%A variation of several orders of magnitude could been seen in the previous measurements
%and the derived upper limits, 
 although great efforts have been made on the study of the exosphere. The 
cold cathode gauge experiments (CCGEs) emplaced on the lunar surface by the Apollo 12, 14, and 15 missions determined
a total concentration of $2\times10^5 \mathrm{cm^{-3}}$ at night, and a weak upper limit of
$2\times10^7 \mathrm{cm^{-3}}$ in daytime (Johnson et al., 1972). The Lunar Atmosphere Composition 
Experiment (LACE) deployed by the final Apollo mission, Apollo 17, reported a firm detection of $\mathrm{^4He}$ in
the night lunar exosphere (e.g., Hodges et al., 1972, 1973).
The upper limits of the concentration of various species, i.e., H, O, C, N, S, Kr, Xe, $\mathrm{H_2}$, and CO, 
were obtained by Feldman \& Morrison (1991) by reanalyzing the Apollo 17 UVS data.
The understanding of the lunar exosphere was much improved by the detection of the $\mathrm{D_2}$-lines of both  
K and Na, which was archieved by ground-based long-slit, high-resolution spectroscopy (e.g., Potter \& Morgan, 1988, 1991; Tyler et al., 1988;
Kolowski et al., 1990; Sprague et al, 1992). The upper limits of the concentration of Si, Al, Ca, Fe, Ti, Ba, and Li were 
obtained by Flynn \& Stern (1996) through ground-based spectroscopy between 3700 and 9700\AA. Stern et al. (1997) 
reported the measured upper limits of the concentration of Mg, $\mathrm{Mg^+}$ and OH radicals by using the HST 
FOS spectroscopy.         
\rm 

%(e.g., Sridharan et al. 2010a,b).

The on-orbit measurements recently done by the Chandra’s Altitude Composition Explore (CHACE)
reported a total pressure of the exosphere of $\sim10^{−7}$
torr (Sridharan et al., 2010a,b), which is higher than
the upper limit previously reported by
the Apollo missions (Heiken et al., 1991) by two orders of magnitude.  CHACE is a mass spectrometer with a mass range of 1-100 amu.   
Its partial pressure sensitivity is $\sim10^{-13}$ torr, and is signiﬁcantly lower than the sensitivity of the LACE by four
orders of magnitude. 
CHACE sampled the exosphere gas
at the sunlit side of the Moon in every four seconds, after
being released from the stationary orbit at an altitude of
about 98 km. The sampling resolution is 0.\symbol{23}1 in 
latitude, and 250m in altitude.
The instrument was separated from the mother spacecraft at 13.3\symbol{23}S and 14\symbol{23}E, and impacted to 
the lunar south pole at $\sim$89\symbol{23}S and -30\symbol{23}W by a oblique trajectory. 
\rm

%The recent on-orbit measurements done by the Chandra’s
%Altitude Composition Explore (CHACE), a mass spectrometer onboard the Moon Impact Probe
%(MIP) of the Indian Chandrayaan-1 mission, reports a total pressure of the exosphere of $\sim10^{−7}$
%torr, which is two orders of magnitude higher than the previously reported upper limits given by
%the Apollo missions (Heiken et al., 1991). In contrast to the results  (dominated by
%$\mathrm{^{20}Ne}$, He, $\mathrm{H_2}$, $\mathrm{^{40}Ar}$, and $\mathrm{CO_2}$) from the Apollo-17 mission (Hodges et al., 1972, 1973), 
The mass spectra taken by CHACE suggest that the exosphere
is dominated by the $\mathrm{H_2O}$ and $\mathrm{CO_2}$ molecules. There is an adjacent small peak at amu = 17 in the mass
spectra, which is likely attributed to the OH radicals. 
 Wang et al. (2011) estimated the surface
concentrations of the OH and $\mathrm{H_2O}$ molecules from the CHACE mass spectra by assuming a hydrostatic equilibrium and an ideal gas law.
By using a scale height of 100km in the estimation, 
the gas temperature was derived from an energy conservation of
each particle in collisionless case. The estimated concentrations are $\sim2\times10^9 \mathrm{cm^{-3}}$ and $\sim6\times10^9 \mathrm{cm^{-3}}$ for the OH and
$\mathrm{H_2O}$ molecules, respectively, which are, however, 
larger than the values predicted by the current models by  
about 6 orders of magnitude.
 \rm

 This paper reports a study on the OH content in the sunshined lunar exosphere based on the 
in-situ measurements taken 
by Lunar-based Ultraviolet Telescope (LUT) (Cao et al., 2011; Wen et al. 2014), a payload of Chinese Chang'e-3 (CE-3) mission (Ip et al., 2014). \rm 
The bandpass of LUT covers the resonant emission lines  $\mathrm{OH(0-0)(A^2\Sigma^+-X^2\Pi)}\lambda3087\AA$, 
which enables 
%The bandpass of LUT covers the OH radical resonant line emission $\mathrm{OH(0-0)(A^2\Sigma^+-X^2\Pi)}$ at around 3087\AA, 
us to study the OH content from the diffuse background level recorded in the images taken by LUT. We refer the 
readers to Wang et al. (2011) for a discussion on the potential effect of the lunar exosphere on 
the sky background detected by LUT.

%unprecedentedly lower than all the previous reports,
%on OH radical (and also MgI) content in the lunar exosphere by the in-situ measurements taken by the
%Lunar-based Ultraviolet Telescope (LUT) (Cao et al. 2011), a payload of the Chinese Chang'e-3 mission. The bandpass of LUT
%covers the OH radical resonant emission $\mathrm{OH(0-0)(A^2\Sigma^+-X^2\Pi)}$ at around 3087\AA, which enables
%us to explore the OH content based on the background level of the images taken by LUT. We refer the 
%readers to Wang et al. (2011) for a discussion on the potential effect of the lunar exosphere on 
%the sky background detected by LUT. 

\section{Lunar-based Ultraviolet Telescope}

%As a payload of Chinese Chang'e-3 mission  (Ip et al., 2014), 
 LUT is the first robotic astronomical telescope
working on the lunar surface in the history of lunar exploration. \rm 
The telescope works in the near-ultraviolet (NUV) band and was developed by National Astronomical Observatories of CAS (NAOC) and 
Xi'an Institute of Optics and Precision Mechanics of CAS (XIOPM). 
We refer the readers to Cao et al. (2011) for more details on the mission's concept and design,  and to Wang et al. (2015) 
for a summary of the prescription in its Table 1. \rm   
Briefly, the optical system of LUT is a F/3.75 Ritchey-Chretien
telescope with an aperture of 150 mm. A pointing flat 
mirror mounted on a two-dimensional gimbal is used to point
and to track a given celestial object. An ultraviolet-enhanced  AIMO \rm CCD
E2V47-20 with a pixel size of 13$\mu$m, manufactured by the e2v Company, is chosen
as the detector mounted at the Nasmyth focus. There are in total $1024\times1024$ active pixels. 
 The detector can be thermal-electrically cooled by as much as 40\symbol{23}C below its environment. 
A NUV coating is applied on the two-lens field corrector located in front of the
focal plan. Two LED lamps with a center wavelength of
286nm are equipped to provide an internal flat field.
\rm
The filed-of-view is $1.36\times1.36$ square degrees.

LUT has been successfully launched on December $2^{\mathrm{nd}}$, 2013 by a Long March-3B rocket, and 
had its first light on December $16^{\mathrm{th}}$, 2013, two days after the CE-3 lander landed on the lunar surface  at 
19.51\symbol{23}W and 44.12\symbol{23}N. With the location on the Moon and the scope of the gimbal, LUT can 
cover the sky region around the north pole of the Moon with a total area about 3600degrees$^2$ (see Figure 2 as an 
illustration). \rm
 By the end of December 2014, LUT has smoothly worked on the Moon for an earth year, and acquired a total of 
more than 130,000 images. Its first six months of operation shows a highly stable photometric performance 
(Wang et al. 2015).
\rm    

Figure 1 shows the measured throughput as a function of wavelength for the whole system.
The throughput peaks at 2500\AA\ with a peak value of $\approx8\%$ and has an effective full width at half maximum (FWHM) of 1080\AA.
The throughput has been determined in the laboratory at pre-launch by a dedicated calibration system. The system is composed of 
a deuterium lamp, a halogen-tungsten lamp, 
a 207D type monochromator from McPherson, Inc., an adjustable diaphragm and two pre-calibrated sensors. The telescope
collects the collimated output light from the monochromator without any intervening optics
element on the light path.

%\begin{table}
%\small
%\caption{The designed performance parameters of LUT}
%\begin{tabular}{lccc}
%\hline
%Parameter& symbol & unit& Value\\
%\hline
%Wavelength range &\dotfill & nm & 245-340 \\
%Aperture size    & $d$ & cm & 15 \\
%F-number              & \dotfill &\dotfill & 3.75\\
%Pixel size of CCD & $d_p$ & $\mu$m &  13 \\
%Average CCD QE  & $\overline{QE}$ & \dotfill &0.4 \\
%Optical efficiency  & $\eta_{\mathrm{opt}}$ & \dotfill & 0.09 \\
%Optics PSF & $f_p$ &\dotfill & 80\% energy within $3\times3$ pixels \\
%CCD Readout noise & $RN$ & $\mathrm{e^-\ pixel^{-1}\ read^{-1}}$ & 8\\
%Dark current & $D$ & $\mathrm{e^-\ s^{-1}\ pixel^{-1}}$ & 1.0 (temperature$<$-20)\\
%CCD Gain &  $G$ & $\mathrm{e^-\ ADU^{-1}}$ & 1\\

%\hline
%\end{tabular}
%\label{table1}
%\end{table}

\section{Observations}

A series of 498 images with extremely low stray light level were obtained by LUT
 when it performed a survey observation and a monitor on RR Lyr type variable star XZ\,Cyg 
($\alpha=\mathrm{19^h:32^m:29^s.3}$, $\delta=\mathrm{+56\symbol{23}:23':17''}$, J2000) in April 10th, 2014 
and June 8th, 2014, respectively. All the images were taken with an exposure time of 30 seconds.
The telescope pointing was fixed with respect to the Moon within each exposure.   
The shift of a star on the focal plane within each exposure
is negligible compared with the size of the point-spread-function (PSF). 
In the monitor, the diurnal apparent motion of the stars was compensated by a step tracking with an interval less than
30 minutes, which enables the star to have a significant shift with respect to the fixed stray light pattern caused 
by the sunshine.
LUT changed its pointing in very 30 minutes in the survey observation for the same reason. 
The survey images used in this paper were taken at a sky region roughly at 
$\alpha=\mathrm{17^h30^m-18^h30^m}$, $\delta=\mathrm{+50-+55\symbol{23}}$. The distribution of all the used pointings is
shown in Figure 2.
\rm 
The elevation of the Sun was
about 12-13\symbol{23} when these images were taken.

\section{Data Reductions}

\subsection{Image reductions}

The raw 498 images were reduced through standard
procedures in astronomy by the IRAF\footnote{IRAF is distributed by
National Optical Astronomy Observatory, which is operated by the Association of Universities for Research in Astronomy, Inc.,
under cooperative agreement with the National Science Foundation.} package, including overscan correction, bias and dark
current subtraction, and flat filed normalization. The bias and dark current images were obtained at the beginning and 
end of each lunar day. All the 498 images and the bias and dark current images were taken 
at a fixed CCD temperature of -40\symbol{23}C.

The instrumental effect removed images were divided into 23
groups according to the pointing of the telescope,  i.e., the pointing is fixed with respect to the Moon 
for each of the groups. The total time of each group was additionally required to be no more than 30 minutes.  
The level and pattern of the stray light caused by the sunshine is therefore believed to be stable for each group, 
%both because of the fixed pointing with respect to the Moon and because of the short time, i.e., close to 30 minutes, 
%spend in each group.
because both pattern and level of the stray light change with time and with telescope pointing with respect to the Moon.  
\rm
The images in each group were then combined to enhance
the signal-to-noise (S/N) ratio of the background. The median value of each pixel was extracted in the 
image combination. One combined image is shown in Figure 3 as an example. 
Each combined image was subsequently smoothed by a box with a size of 5$\times$5 pixels by using the IRAF package to
further enhance its background S/N ratio. 
%further enhanced by a smoothing with
%The S/N ratio is further enhanced by smoothing each of the combined images by

\subsection{Pixel distributions}

 The aim of this study is to estimate the content of the sunshined lunar exosphere from the 
diffuse sky background caused by the resonant line emitted from the species in the exosphere.    
This task is, however, complicated by several issues. At first,  
%In addition to a diffuse contribution to the sky background, 
although the stray light level is extremely low in the 498 images, 
a stray light pattern can still be clearly identified in  
all these images (see Figure 3 as an example again), which results in a nonuniform 
contamination on the potential diffuse background caused by the resonant line emission.  
Secondly, both pattern and level of the stray light vary with both time and pointing with respect to the Moon.  
Finally, the contamination caused by the stray 
light can not be entirely excluded from the images. 
%even though the structured stray light can be easily identified in the images. 

\rm

To address the first issue,
we divided each combined and smoothed image into 16 sub-images each with a physical size of 200$\times$200
pixels (see an illustration of the grid and the identification of each of the sub-images in Figure 3).
A pixel distribution was extracted from each sub-image.  
%mininize the pollution due to the stray light,  
%the pixel distributions were extracted from 16 sub-images each with a physical size of 200$\times$200
%pixels for each combined and smoothed image (see the grid and the identification of each of the sub-images in Figure 2). 
Generally speaking, each distribution shows a strong peak superposed by a secondary peak or a tail at its large count-rate end.
Our examination by eyes indicated that both the secondary
peak and the large count-rate wing are likely resulted from the
pollution caused by the stray light pattern. 
We then fitted each of the distributions by a sum of multiple Gaussian functions.  The fitting is schemed in 
Figure 3 for two typical distributions extracted from different sub-images (i.e., different sub-image identifications)
that were taken at different time with different pointings with respect to the Moon. 
%The lower panel in the figure shows the pixel distribution extracted from sub-image M of 
%combined image No. 2 taken in April 10th 2014, and the upper panel the distribution extracted from sub-image N of combined image No. 17 taken in June 8th 2014.
The diffuse background level of each sub-image was determined from the fitted expected value of the Gaussian function that 
reproduces the strong peak of the distribution.

\rm

%The difference between the two distributions can be explained by a variation of the contamination of the stray light.
%Both the pattern and level of the stray light changes with both pointing with respect to the Moon and time.    
%The diffuse background level of each sub-image was determined from the fitted expected value of the Guassian function that 
%reproduces the strong peak of the distribution. One should be bear in mind that a diffuse contribution by the stray 
%light can not be entirely excluded from the diffuse background level determined here. In a given combined image, 
%with the 16 determined diffuse background levels at different CCD regions, 
%the lowest one was used as an assessment of the diffuse background level with mininized contamination by the stray light. 

In a given combined image, its underlying diffuse background level with minimized stray light contamination 
was extracted from the 16 diffuse background levels determined at different CCD regions by the lowest one. 
Table 1 lists these values for the 23 combined and smoothed images.
% The determined diffuse background levels are tabulated in Table 1 for the 23 combined and smoothed images. 
For each of the combined images,
the Column (3) lists the count rate of its diffuse background along with the corresponding uncertianty at 1$\sigma$ significance level. 
The reported count rate was converted from the originally recorded ADU by using a measured CCD gain of 1.59. The uncertainty 
was obtained from the fitted width of the Gaussian function\footnote{The formal uncertainty of 
the expected value of the fitted Gaussian function is $\sim10^{-4}\mathrm{e^-\ s^{-1}\ pixel^{-1}}$, which is much smaller than the errors reported 
in Table 1.}. The corresponding S/N ratio and 
the identification of the sub-image in which the diffuse background level was measured are tabulated in Column (5) and (2), 
respectively. One can see from the Column (5) that the lowest stray light contamination always occurs
at the two left corners of the CCD in all the 23 combined images,
which is in agree with our inspection by eyes.  

As shown by the Column (3) in Table 1, one can see a considerable variation of the determined diffuse background count rates. The variation
can be easily understood because these combined images were taken at different time with different pointings with respect to the Moon. 
In order to further minimize the stray light contamination, the lowest diffuse background level 
with a count rate of $\mathrm{0.096\pm0.024 e^-\ s^{-1}\ pixel^{-1}}$ (with a S/N ratio of 3.96) is adopted 
in the subsequent calculations. 
The value was derived from the sub-image B in the combined image No. 31\footnote{Two comparable 
values can be found in the sub-image M in the combined images No. 28 and 29.}. 

Finally, we emphasize that this adopted value should be used as an upper limit of the diffuse background level 
caused by the resonant line emission, because the contamination due to the stray light can not be 
fully removed from the observed background.

\rm

%Comparing the 23 extracted diffuse background levels finally enables us to identify the lowest background level
%with an average count rate of $\mathrm{0.096\pm0.007 e^-\ s^{-1}\ pixel^{-1}}$ at a S/N ratio of 3.96, where the error 
%corresponds to a 1$\sigma$ significance level and a measured CCD gain of 1.59 is used. 

\section{Transformation from the Background to Atmosphere Content}

One should be bear in mind that the lowest diffuse background count rate of $\mathrm{0.096\pm0.024 e^-\ s^{-1}\ pixel^{-1}}$ derived above
contains the contributions from: 1) the stray light 
caused by the sunshine; 2) the zodiacal light from the solar system, and 3) the emission from the resonant transitions in the lunar exosphere.
These complications result in the fact that only an upper limit of the content of a given species can be inferred from the 
LUT's in-situ measurements. 
%The diffuse contribution to the background due to the stray light can not be removed 
%by our fittings to the extracted pixel distributions.  
The potential transitions within the LUT's wavelength coverage includes the $\mathrm{OH(0-0)(A^2\Sigma^+-X^2\Pi)}$ at 
3087\AA, MgI$\lambda$2853 and AlI$\lambda3092$ (Morgan \& Killen, 1997, see below).

\subsection{Contribution from the zodiacal light}

We estimated the underlying contribution from the zodiacal light in the LUT bandpass through
a combination of the zodiacal map, the zodiacal spectral shape (both taken from Leinert et al., 1998) and the throughput of
LUT. The count rate due to the zodiacal light can be calculated through the integration of
\begin{equation}
 R_{\mathrm{ZL}}=\frac{1}{4}\pi d^2\Delta\Omega\int I_\lambda^{\mathrm{ZL}}(\beta-\beta_\odot,\lambda')S_\lambda d\lambda
\end{equation}  
where $d=150$mm is the diameter of the aperture of LUT, $S_\lambda$ the system throughput at different wavelengths, and 
$\Delta\Omega=22.7\mathrm{arcsec^2\ pixel^{-1}}$ the solid angle of each pixel.
$I_\lambda^{\mathrm{ZL}}(\beta-\beta_\odot,\lambda')$ is the spectrum of the surface brightness of the zodiacal light measured at the Earth, 
 whose shape and absolute level are a function of the viewing direction ($\beta-\beta_\odot$, $\lambda'$) with respect to the Sun.  
A fixed spectral shape that is best measured at the ecliptic equator with a solar elongation $\beta-\beta_\odot=90$\symbol{23} was adopted in our 
integration. The absolute intensity of the spectrum was determined from the zodiacal map
by a 2-dimensional interpolation.  
% according to the 
%calculated viewing direction. 
With the calculated viewing direction of $\beta-\beta_\odot\sim110\symbol{23}$ and $\lambda'\sim 75\symbol{23}$  (see Figure 2 as an illustration), \rm
the count rate on the focal plane caused by the zodiacal light was finally estimated to be $ R_{\mathrm{ZL}}\sim0.02\mathrm{e^-\ pixel^{-1}\ s^{-1}}$.

%The integration finally yields a count-rate of $\sim0.02\mathrm{e^-\ pixel^{-1}\ s^{-1}}$ caused by the zodiacal light, which results 
%in an improved upper limit of $R\sim0.07\mathrm{e^-\ pixel^{-1}\ s^{-1}}$.    

%With the LUT pointing direction with respect to the Sun, the zodiacal map
%yields a count-rate of $\sim0.02\mathrm{e^-\ pixel^{-1}\ s^{-1}}$ on the focal plane of LUT through 
%the integration of 
%\begin{equation}
% R_{\mathrm{ZL}}=\frac{1}{4}\pi d^2\Delta\Omega\int I_\lambda^{\mathrm{ZL}}(\beta-\beta_\odot,\lambda')S_\lambda d\lambda
%\end{equation}
%where $d=150$mm is the diameter of the aperture of LUT, $S_\lambda$ the system throughput at different wavelengths, and 
%$\Delta\Omega=22.7\mathrm{arcsec^2\ pixel^{-1}}$ the solid angle of each pixel. 
%$I_\lambda^{\mathrm{ZL}}(\beta-\beta_\odot,\lambda')$ is the surface brightness of the zodiacal light derived from
%the zodiacal light map observed from the Earth, where $\beta-\beta_\odot$ and $\lambda'$ are the
%solar elongation in equator and latitude in ecliptic, respectively.
%The exclusion of the calculated contribution from the zodiacal light yields a background
%level of $R\sim0.07\mathrm{e^-\ pixel^{-1}\ s^{-1}}$.

\subsection{Lunar exosphere content}

\subsubsection{Method}

 A corrected diffuse background count rate of $R\sim0.08\pm0.02\mathrm{e^-\ s^{-1}\ pixel^{-1}}$ was finally 
derived after the exclusion of the estimated contribution due to the zodiacal light.
%After removing the estimated contribution due to the zodiacal light from the determined minimum diffuse background level,
By assuming this background count rate is entirely caused by the resonant scattering line emission (within the LUT bandpass)
of a given species in the lunar exosphere, an upper limit of the column density $N$ of the species can be
calculated as follows.

\rm

At the beginning, the diffuse background count rate $R$ was 
transformed to the resonant line intensity $I_s$ as 
\begin{equation}
 R=\frac{1}{4}\pi d^2S_\lambda\Delta\Omega I_s
\end{equation}
where $S_\lambda$ is the measured throughput of LUT at the line wavelength of $\lambda$.
The calculated line intensity was subsequently used to estimate the  
column density $N$ of the species in the optical thin case (Flynn \& Stern, 1996):
\begin{equation}
 4\pi I_s\cos\theta = gN
\end{equation}
where $\theta\sim37\symbol{23}$ is a parameter accounting for the
angular distance from the pointing of LUT to the local zenith, and
$g$ the corresponding fluorescence efficiency at 1AU
(i.e., the g-factor). 
The g-factor is defined as an emission probability per atom in units of 
$\mathrm{photon\ s^{-1}\ atom^{-1}}$, and is determined by summing the probabilities of all transitions from multiple states 
whose population partitions are solved from the detailed equilibrium of every state that 
is usually not in thermodynamic equilibrium.

\subsubsection{OH radicals}

Taking the g-factor of $g=2.3\times10^{-4}\ \mathrm{photons\ s^{-1}\ molecule^{-1}}$ (Schleicher et al., 1988) 
at 1AU for the OH 3087\AA\ transitions, the zodical light-corrected diffuse background count rate of $R\sim0.08\pm0.02\mathrm{e^-\ pixel^{-1}\ s^{-1}}$ was transferred to
an upper limit of the column density of $N\sim(6.0\pm1.5)\times10^{11}\ \mathrm{cm^{-2}}$ for the OH radicals. 
A slightly lower upper
limit of $N\sim(4.0\pm1.0)\times10^{11}\ \mathrm{cm^{-2}}$ can be obtained if the g-factor at a temprature of 200K of
$g=3.5\times10^{-4}\ \mathrm{photons\ s^{-1}\ molecule^{-1}}$ (Stevens et al., 1999) was adopted in the calculation.

We transformed the calculated column density to the surface concentration
$n_0$ as $N=n_0\Delta H$ by assuming a hydrostatic equilibrium, where $\Delta H$ is the model dependent scale height of the vertical
density profile. The light OH radicals are believed to be thermal released (Stern, 1999).
By adopting a typical scale height of $\Delta H=100 \mathrm{km}$ (Wurz et al., 2007), two comparable surface concentrations of
 $\sim(6.0\pm1.5)\times10^4\ \mathrm{cm^{-3}}$ and $\sim(4.0\pm1.0)\times10^4\ \mathrm{cm^{-3}}$ can be obtained from the two estimated 
column densities, which depends on the adopted g-factor values.  The slight difference between the two adopted g-factor values
enables us to report an 
upper limit of the surface OH concentration of $\sim10^4\ \mathrm{cm^{-3}}$, which    
\rm
is the best constraint on the OH content in the lunar exosphere to date (see a comparison in the next section). 
 The calculated results are summarized in Table 2. \rm 
The total mass of the OH radicals in the exosphere was estimated to be no more
than $\sim1.5\times10^3\ \mathrm{kg}$.

\subsubsection{Neutral magnesium and aluminum}

Besides the OH resonant emission at 3087\AA, the LUT bandpass
covers two strong resonant emission lines, they are MgI$\lambda$2853 and AlI$\lambda$3092 (Morgan \& Killen, 1997). We calculated the
upper limits of both column density and surface concentration for the two species by the same method described in Section 5.2.1.

 Using the g-factor value of the MgI$\lambda$2853 resonant line emission of $2.3\times10^{-4}\ \mathrm{photons\ s^{-1}\ atom^{-1}}$ reported in
Morgan \& Killen (1997), we derived an upper limit of the column density of the neutral magnesium of $\sim(2.0\pm0.5)\times10^{9} \ \mathrm{cm^{-2}}$ 
from the zodiacal light-corrected diffuse background count rate of $R\sim0.08\pm0.02\mathrm{e^-\ pixel^{-1}\ s^{-1}}$. The assumption of a hydrostatic equilibrium
yielded an upper limit of the surface concentration of $\sim(2.0\pm0.5)\times10^{2} \ \mathrm{cm^{-2}}$ when a scale height of 100km is 
adopted.  
\rm
Ion sputter model shows that the heavy atoms like Mg and Al can reach a considerably large scale
height of $\sim1,000\ \mathrm{km}$ due to the sputtering by the solar wind (Wurz et al., 2007). If so, 
the inferred surface concentration is $< 20\pm5\ \mathrm{cm^{-2}}$ for the neutral magnesium. 
It is noted that both upper limits of the surface concentration obtained with different scale height values 
are lower than those reported in the previous studies (see a comparison in the next section).

 An upper limit of the column density of the neutral aluminum was calculated to be $\sim(4.0\pm1.0)\times10^{9} \ \mathrm{cm^{-2}}$ based on the 
g-factor of the AlI$\lambda$3092 resonant line emission of $3.8\times10^{-2}\ \mathrm{photons\ s^{-1}\ atom^{-1}}$ (Morgan \& Killen 1997). 
The inferred surface concentration is $<(4.0\pm1.0)\times10^{2} \ \mathrm{cm^{-3}}$ for a scale height of 100km, and 
 $<40\pm10 \ \mathrm{cm^{-3}}$ for a scale height of 1,000km. 
\rm

 All these calculated results are tabulated in Table 2. \rm Note that the surface
concentrations listed in the table are all based on a scale height of 100 km.

%The results are listed in Table 1. Again, we emphasize that the obtained upper
%limits are lower than that reported inall the  previous studies, except for AlI (see below). . Ion
%sputter model shows that the heavy atoms like Mg and Al could reach a considerably large scale
%height of $\sim1000\ \mathrm{km}$ due to the sputtering by the solar wind (Wurz et al., 2007). If so, the upper limits of the surface
%concentration reported in Table 1 are overestimated by an order of magnitude.

\section{Comparisons}

The extremely low density of the lunar exosphere results in a considerable challenge in determining
its content. The current study shows that the LUT's in-situ measurements provide the best constraints
on the content of the lunar exosphere to date, both because of the high efficiency of LUT and 
because of the avoidance of the contamination caused by the instrument itself. 
%The obtained
%upper limits are at least one to several orders of magnitude lower than the previous claims, except
%for AlI. 

 The upper limit of $<10^4\ \mathrm{cm^{-3}}$ derived for the OH radicals is 
lower than that derived from the HST low resolution spectroscopy (i.e, $<10^6\mathrm{cm^{-3}}$, Stern et al., 1997) by
about two orders of magnitude, and is lower than 
that inferred from the mass spectra taken by the Chandrayaan-1 mission (see Introduction and Wang et al. (2011) for the details)
by about 6 orders of magnitude.

\rm
%derived from the HST low resolution spectroscopy (i.e, $<10^6\mathrm{cm^{-3}}$, Stern et al., 1997), and about 6 orders of magnitude lower than
%that inferred from the mass spectra taken by Chandrayaan-1 mission (see Introduction and Wang et al. (2011) for the details.). 

 We argue that the upper limit derived from the LUT's in-situ measurements is very close to   \rm
%The inferred upper limit on the surface concentration of the OH radical is roughly
%consistent with 
the prediction of the ion sputtering model. The chemically adsorbed $\mathrm{H_2O/OH}$
molecules (Stern, 1999; Starukhina et al., 2000; Arnold, 1979) could be sputtered by the solar wind protons, which
produces water vapor and gaseous OH radicals in the exosphere through chemical reactions (e.g., Morgan et al., 1997; Wurz et al., 2007; 
Killen \& Ip, 1999; Hunten et al., 1997; Johnson et al., 1991; Killen et al., 1999). The sputtering results
in a surface $\mathrm{H_2O/OH}$ concentration $n_0\sim f_a j\eta\upsilon^{-1}\sim10^3\ \mathrm{cm^{-3}}$,
where $j\sim10^8\ \mathrm{p^+\ cm^{-2}\ s^{-1}}$ is the
proton flux of the solar wind, $\eta\sim0.1$ the production rate per proton (Crider \& Vondrak, 2002), $\upsilon\sim10^2\ \mathrm{m\ s^{-1}}$ the
typical velocity of a particle,  and $f_a\sim0.8$ the fraction of the solar protons that are absorbed at the lunar surface. 
Recent observations carried out by Chandrayaan-1/SARA (Sub-kev Atom Reflection Analyzer) found that about $\sim20\%$ of the incident solar protons
is backscattered to space from the surface (Bhardwaj et al., 2012). The backscattered proton becomes a hydrogen atom by recombining with an electron.     

\rm

%The same situation occurs for MgI.
Previous study reported an upper limit of $<6,000\ \mathrm{cm^{-3}}$ at a significance level of 5$\sigma$ for 
the neutral magnesium (Stern et al., 1997),  which
is larger than the value reported in this study by 
about two (for $\Delta H=100\mathrm{km}$) or three (for $\Delta H=1000\mathrm{km}$) orders of magnitude. 
\rm
HST UV spectroscopy observation reported a 5$\sigma$ upper limit of $<11,000\mathrm{cm^{-3}}$ for neutral aluminum.  This value is
larger than that given by the LUT's measurements by about two orders of magnitude, when 
a scale height of 100km was adopted in our calculations.
\rm
A much lower upper limit of $< 55\mathrm{cm^{-3}}$ at a significance level of 5$\sigma$ was obtained from the ground-based spectroscopy (Flynn \& Stern, 1996),
which is very close to the upper limit derived here when a scale height of $\Delta H=1000\mathrm{km}$ was used.

\section{Conclusions}

An unprecedented upper limit of the OH (MgI) content in the lunar exosphere is 
obtained from the in-situ measurements carried out by LUT. 
The inferred column density 
and surface concentration are $<10^{11}\ \mathrm{cm^{-2}}$  and $<10^{4}\ \mathrm{cm^{-3}}$ for the OH radicals, respectively, 
which is close to the value predicted by the sputtering model. 
In addition, a surface concentration of $<10^{2}\ \mathrm{cm^{-3}}$ is obtained for neutral magnesium 
from the same measurements and method.

\section*{Acknowledgments}
We thank the anonymous referee for his/her helpful suggestions for improving the manuscript.
The authors thank the outstanding work of the LUT team and the support by the
team of the ground system of Chang'e-3 mission. The study is supported by the Key Research Program of
Chinese Academy of Science (KGED-EW-603) and by the National Basic Research Program of China
(973-program, Grant No. 2014CB845800). JW is supported by the National Science Foundation of
China under Grant 11473036. MXM is supported by the National Science Foundation of China under
Grant 11203033.

%\section{acknowledgements}

%\section{Citations}
%\label{Section 3}

%\begin{itemize}
%\item Parenthetical: \verb|\citep{WB96}| produces \citep{WB96}.
%\item Textual: \verb|\citet{WB96}| produces \citet{WB96}.
%\item An affix and part of a reference:
%   \verb|\citep[e.g.][Ch. 2]{WB96}|
%   produces \citep[e.g.][Ch. 2]{WB96}.
%\end{itemize}

%%%%%%%%%%%%%%%%%%%%%%%%%%%%%%%%%%%%%%%%%%%%%%%%%%%%%%%%%%%%%%%%%%%%%%%%%%%%%
%% Appendices
% \appendix

\clearpage

\begin{figure}
%\epsscale{1.0}
\begin{center}
\includegraphics[height=5cm]{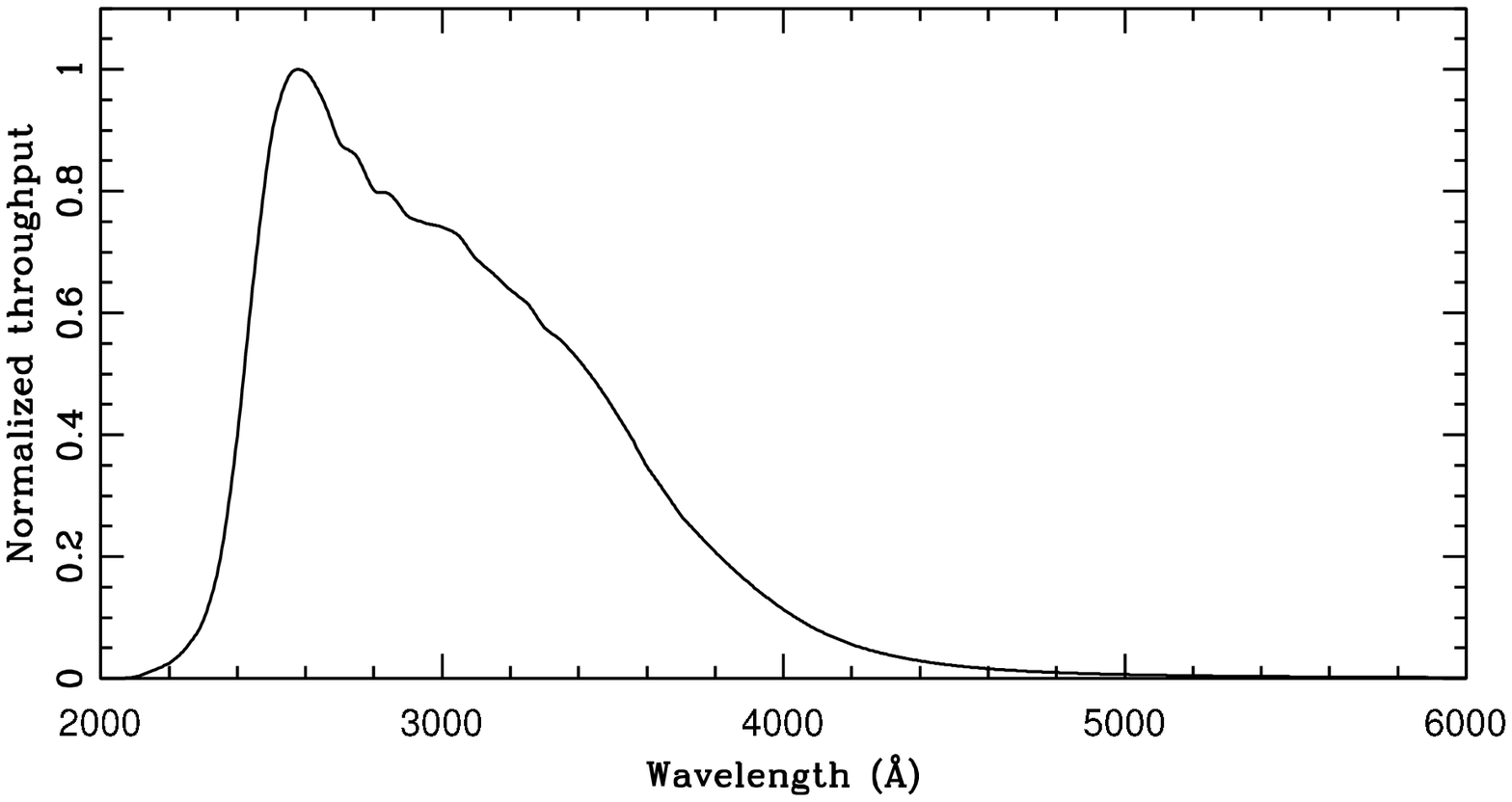}
\caption{The normalized throughput of LUT plotted against wavelength.}
\end{center}
\end{figure}

\begin{figure}
%\epsscale{1.0}
\begin{center}
\includegraphics[height=5cm]{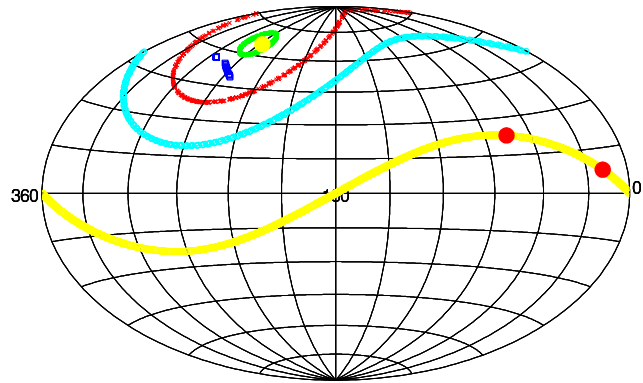}
\caption{The distribution of the pointings used in this paper is shown in the J2000 mean equator coordinate system by the blue squares. 
The LUT's available sky is located between the red and green circles. The locus of
the zenith of LUT is shown by the cyan curve. The large yellow point marks the north pole of
the Moon. The solid yellow line shows the equator in ecliptic. The locations of the Sun in April 10th, 2014 and June 8th, 2014 
are marked by the large red points.
}
\end{center}
\end{figure}

\begin{figure}
%\epsscale{1.0}
\begin{center}
\includegraphics[height=10cm]{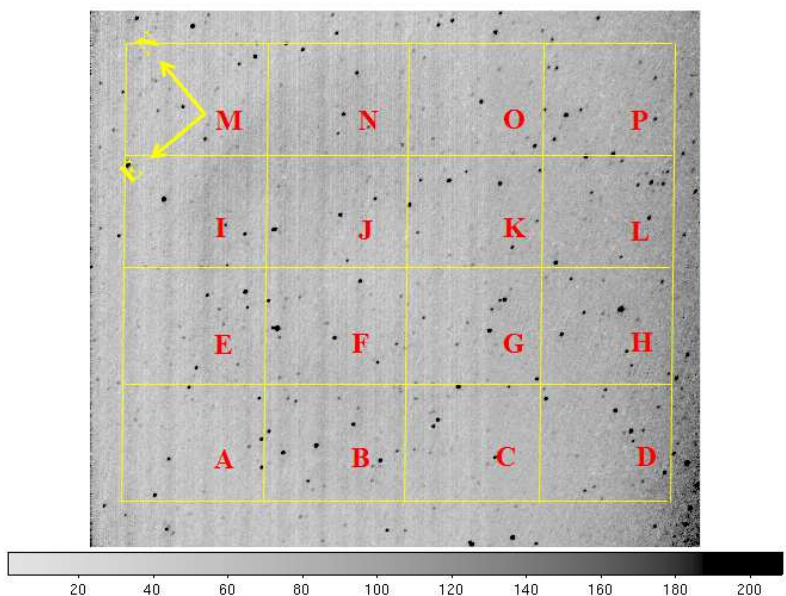}
\caption{An example of the reduced and combined image.  The combined image was taken in the survey observation in 
April 10th, 2014. The center of the image is at $\alpha=\mathrm{18^h:00^m:08^s.6}$, $\delta=\mathrm{+51\symbol{23}:22':09''}$ (J2000).
The size of the image is about $1.35\times1.35\mathrm{degrees^2}$. The east and  north directions are marked by the two yellow
arrows labeled at the left-upper corner.   
The points in the image are stars, and the strips at the left side are due to the CCD electronics. The bar underneath the 
image shows the scale of variation of brightness in logarithm. \rm   
A pattern due to the stray light
caused by the sunshine can be identified from a ring around the image center and from a gradual brightening
at the CCD right fringe.  The horizontal and vertical yellow lines illustrate the grid used to separate the image into 
16 sub-images each with a pixel size of $200\times200$. \rm }
\end{center}
\end{figure}

\begin{figure}
%\epsscale{1.0}
\begin{center}
\includegraphics[height=8cm]{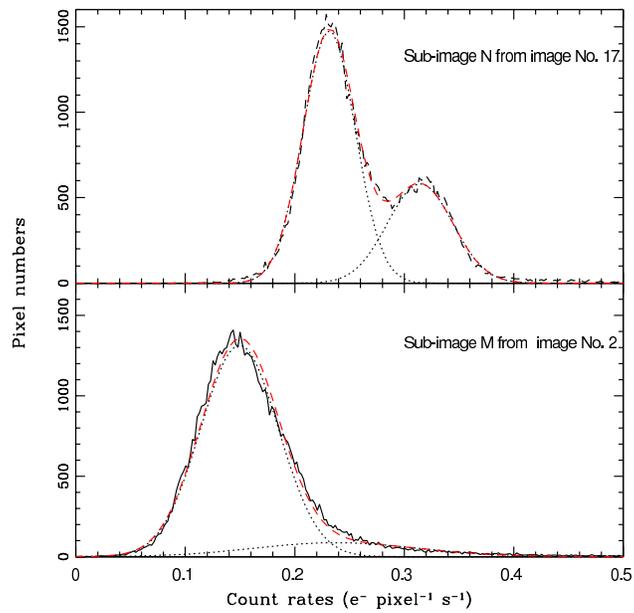}
\caption{Examples of fitting an extracted pixel distribution by a sum of two Gaussian functions.  
In each panel, the observed distribution and the best fit curve are plotted by the black solid and
red dashed lines, respectively. The two fitted Gaussian functions are shown by the dotted lines. }
\end{center}
\end{figure}

\clearpage

\begin{table}
\centering
\caption{The determined underlying background count rates of the 23 combined images.}
\medskip
\scriptsize
\begin{tabular}{ccccc}
\hline
%Group number & sub-image ID & count-rate &  error in counter-rate & S/N\\
%             &              & $\mathrm{e^{-}s^{-1}pixel^{-1}}$ &  $\mathrm{\times10^{-4} e^{-}s^{-1}pixel^{-1}}$  & \\
Combined image ID & sub-image ID & count-rate &   S/N\\
             &              & $\mathrm{e^{-}s^{-1}pixel^{-1}}$   & \\
(1) & (2) & (3) &(4) \\ 
\hline
02 & M & $0.150\pm0.036$   &    4.21 \\
03 & M & $0.195\pm0.032$   &    6.07 \\
04 & M & $0.201\pm0.032$   &    6.28 \\
07 & M & $0.169\pm0.030$   &    5.64 \\
08 & M & $0.154\pm0.029$   &    5.35 \\ 
09 & M & $0.136\pm0.028$   &    4.78 \\
12 & M & $0.276\pm0.029$   &    9.53 \\
13 & M & $0.391\pm0.030$   &    13.25 \\
15 & N & $0.291\pm0.026$   &    11.20 \\
16 & M & $0.350\pm0.026$   &    13.59 \\
17 & N & $0.231\pm0.024$   &     9.66 \\
18 & N & $0.172\pm0.022$   &     7.91 \\
20 & N & $0.224\pm0.022$   &    10.15 \\
21 & N & $0.170\pm0.019$   &     8.98 \\
22 & N & $0.172\pm0.020$   &     8.54 \\
23 & M & $0.191\pm0.027$   &     6.98 \\
24 & M & $0.182\pm0.027$   &     6.69 \\
25 & M & $0.144\pm0.025$   &     5.76 \\
26 & M & $0.158\pm0.027$   &     5.83 \\
28 & M & $0.106\pm0.027$   &     3.91 \\
29 & M & $0.118\pm0.029$   &     4.10 \\
31 & B & $0.096\pm0.024$   &     3.96 \\
32 & E & $0.164\pm0.032$   &     5.04 \\

\hline
%\hline
\end{tabular}
\end{table}

%\begin{table}
%\centering
%\caption{The fitted lowest background levels in 23 combined images}
%\medskip
%\scriptsize
%\begin{tabular}{ccccc}
%\hline
%r & sub-image ID & count-rate &  error in counter-rate & S/N\\
%             &              & $\mathrm{e^{-}s^{-1}pixel^{-1}}$ &  $\mathrm{\times10^{-4} e^{-}s^{-1}pixel^{-1}}$  & \\
%(1) & (2) & (3) &(4) & (5) \\ 
%\hline
%02 & M & 0.150   &   6.7  &    4.21 \\
%03 & M & 0.195   &   5.3  &    6.07 \\
%04 & M & 0.201   &   6.3  &    6.28 \\
%07 & M & 0.169   &   4.3  &    5.64 \\
%08 & M & 0.154   &   6.6  &    5.35 \\ 
%09 & M & 0.136 &   7.0  &    4.78 \\
%12 & M & 0.276 &   3.1  &    9.53 \\
%13 & M & 0.391  &   4.3  &    13.25 \\
%15 & N & 0.291 &   5.2  &    11.20 \\
%16 & M & 0.350  &   5.0  &    13.59 \\
%17 & N & 0.231  &   2.3  &    9.66 \\
%18 & N & 0.172  &   4.5  &    7.91 \\
%20 & N & 0.224 &   5.7  &    10.15 \\
%21 & N & 0.170 &   5.6  &    8.98 \\
%22 & N & 0.172 &   8.5  &    8.54 \\
%23 & M & 0.191  &   6.4  &    6.98 \\
%24 & M & 0.182 &   5.6  &    6.69 \\
%25 & M & 0.144 &   3.0  &    5.76 \\
%26 & M & 0.158  &   4.0  &    5.83 \\
%28 & M & 0.106 &   3.6  &    3.91 \\
%29 & M & 0.118 &   3.2  &    4.10 \\
%31 & N & 0.096 &   2.7  &    3.96 \\
%32 & M & 0.164 &   6.7  &    5.04 \\

%\hline

%\hline
%\end{tabular}
%\end{table}

\begin{table}
\centering
\caption{Calculated column densities and surface concentrations for different species.}
\medskip
\scriptsize
\begin{tabular}{cccccc}
\hline
Species & Wavelength & N & $n_0$ & g-factor @ 1AU & Reference\\
        &  \AA      & $\mathrm{cm^{-2}}$ & $\mathrm{cm^{-3}}$ & $\mathrm{photons\ s^{-1}\ molecule^{-1}}$ & \\
(1) & (2) & (3) &(4) & (5) & (6)\\ 
        
\hline
%MgI & 2853 & $2.4\times10^9$  & $2.4\times10^2$ & $5.4\times10^{-2}$ & Morgan \& Killen 1997\\
%OH  & 3087 & $5.9\times10^{11}$ & $5.9\times10^4$ & $2.3\times10^{-4}$ & Schleicher et al. 1988 \\
%    &      & $3.8\times10^{11}$ & $3.8\times10^4$ & $3.5\times10^{-4}$ & Stevens et al. 1999\\
%AlI & 3092 & $3.6\times10^9$  & $3.6\times10^2$ & $3.8\times10^{-2}$ & Morgan \& Killen 1999\\
MgI & 2853 & $(2.0\pm0.5)\times10^9$  & $(2.0\pm0.5)\times10^2$ & $5.4\times10^{-2}$ & Morgan \& Killen (1997)\\
OH  & 3087 & $(6.0\pm1.5)\times10^{11}$ & $(6.0\pm1.5)\times10^4$ & $2.3\times10^{-4}$ & Schleicher et al. (1988) \\
    &      & $(4.0\pm1.0)\times10^{11}$ & $(4.0\pm1.0)\times10^4$ & $3.5\times10^{-4}$ & Stevens et al. (1999)\\
AlI & 3092 & $(4.0\pm1.0)\times10^9$  & $(4.0\pm1.0)\times10^2$ & $3.8\times10^{-2}$ & Morgan \& Killen (1997)\\

\hline
\end{tabular}
\end{table}

\end{document}